\documentclass[prb,preprint,amsmath]{revtex4}
\pdfoutput=1
\usepackage{amsmath}
\usepackage{units}
\usepackage{graphicx}
\usepackage{braket}
\usepackage{xfrac}
\usepackage{enumitem}
\usepackage[normalem]{ulem}
\usepackage{booktabs}
\usepackage{tabularx}
\usepackage{multirow}
\usepackage{hhline}
\usepackage{physics}
\usepackage{etoolbox}

\usepackage[sort&compress]{natbib}
\renewcommand{\op}[1]{\ensuremath{\hat{#1}}}
\newcommand{\inlinecomment}[1]{}

\newcolumntype{Y}{>{\centering\arraybackslash}X}

\begin{document}

\title {Improving performance in quantum mechanics with explicit incentives to correct mistakes}

\pacs {1}

\author{Benjamin R. Brown}
\author{Andrew Mason}
\author{Chandralekha Singh}
\affiliation{Department of Physics and Astronomy, University of Pittsburgh, Pittsburgh, PA 15260, University of Central Arkansas, Conway, AR  72035}

\begin{abstract}
An earlier investigation found that the performance of advanced students in a quantum mechanics course did not automatically improve from midterm to final exam on identical problems even when they were provided the correct solutions and their own graded exams. Here, we describe a study, which extended over four years, in which upper-level undergraduate students in a quantum physics course were given four identical problems in both the midterm exam and final exam.  Approximately half of the students were given explicit incentives to correct their mistakes in the midterm exam. In particular, they could get back up to 50\% of the points lost on each midterm exam problem. The solutions to the midterm exam problems were provided to all students in both groups but those who corrected their mistakes were provided the solution after they submitted their corrections to the instructor.  The performance on the same problems on the final exam suggests that students who were given incentives to correct their mistakes significantly outperformed those who were not given an incentive. The incentive to correct the mistakes had greater impact on the final exam performance of students who had not performed well on the midterm exam. 
\end{abstract}

\maketitle
\section{Introduction}

Helping students learn to think like a scientist is an important goal of most courses for science and engineering majors at all levels \cite{heller1984prescribing,larkin1979understanding,larkin1981cognition,Maloney1994,reif1981teaching,reif1995millikan,reif2008scientific,van1991learning}. Meeting this goal is also critical to prepare an additional one million STEM professionals in ten years according to the recommendations of the United States President's Council of Advisors on Science and Technology (PCAST) report\cite{pcast}. In order to achieve this goal, it may be beneficial if students are provided incentives to help them learn to think like 
scientists.

One attribute of experts is that they are likely to use problem solving as an opportunity for learning \cite{larkin1979understanding,larkin1981cognition,Maloney1994,reif1981teaching,reif1995millikan,reif2008scientific}. In particular, experts automatically reflect upon their mistakes in their problem solutions in order to repair, extend and organize their knowledge structure.   Since experts learn from their own mistakes, they are unlikely to make the same mistakes when asked to solve a problem a second time, especially if they have had access to a correct solution. Unfortunately, for many students in physics courses, problem solving is a missed learning opportunity \cite{mason2008identifying,cohen2008identifying,mason2008identifying,singh2007physicsII,yerushalmi2007physics,yerushalmi2008effect,yerushalmi2012atypical,yerushalmi2012students,henderson,mason2010advanced}. 
  Without guidance, students often do not reflect upon the problem solving process after solving problems in order to learn from them nor do they make an effort to learn from their mistakes after the graded problems are returned to them. 
 
However, closing the ``performance gap'' between high and low achieving students and ensuring that all students excel in science courses are important goals of science education research\cite{science}. Prior research in introductory physics suggests that instruction can explicitly prompt students to learn successfully from their mistakes by rewarding them for correcting their mistakes.
\cite{mason2008identifying,cohen2008identifying,mason2008identifying,singh2007physicsII,yerushalmi2007physics,yerushalmi2008effect,yerushalmi2012atypical,yerushalmi2012students,henderson}.  This type of activity, if it is repeated in many courses throughout the undergraduate course work, also has the potential to help students learn to make use of problem solving as a learning opportunity.

Prior research also suggests that only providing students worked examples is insufficient \cite{atkinson} and effective approaches to learning involve students engaged in meta-cognition or self-monitoring while they solve problems \cite{meta,meta2,chi1989self}. For example, research suggests that students who went through a productive failure cycle, in which they worked in groups to solve complex ill-structured math problems without any scaffolding support struggled to learn up until a consolidation lecture by the instructor \cite{kapur}. However, these students from the productive failure condition significantly outperformed their counterparts from the lecture and practice condition  on both well- and ill-structured problems on the posttests \cite{kapur}. After the posttest, they also demonstrated significantly better performance in using structured-response scaffolds to solve problems on a new topic not even covered during instruction. Similarly, Schwartz et al. have proposed invention tasks to prepare students for future learning \cite{schwartz}.

One characteristic of prior research studies has been that they have mostly focused on how introductory physics students differ from physics experts \cite{chi1981categorization,dufresne2005knowledge,hardiman1989relation,larkin1980expert,singh2002physical} and strategies that may help introductory students learn to learn \cite{elby2001helping,eylon1984effects,meltzer2005relation,ozimek2005retention,van1991overview,yerushalmi2012atypical,yerushalmi2012students}. By comparison, few investigations have focused on the learning skills of advanced physics students, although some investigations have been carried out on the difficulties advanced students have with advanced topics such as quantum physics and how to help them learn quantum mechanics better \cite{bao2002understanding,belloni2007open,singh2001student,singh2008student,wittmann2002investigating}.

It is often implicitly assumed that, unlike students in introductory physics courses, most students who have made it through an entire undergraduate physics curriculum have not only learned a wide body of physics content but have also picked up the habits of mind and self-monitoring skills needed to build a robust knowledge structure \cite{lin}. Many physics instructors take for granted that advanced physics students will learn from their own mistakes in problem solving without explicit prompting, especially if they are given access to clear solutions. They often assume that, unlike introductory students, advanced students have become independent learners and they will take the time out to learn from their mistakes, even if the instructors do not reward them for fixing their mistakes, e.g., by explicitly asking them to turn in, for course credit, a summary of the mistakes they made and writing down how those mistakes can be corrected \cite{mason2008identifying,cohen2008identifying,mason2008identifying,singh2007physicsII,yerushalmi2007physics,yerushalmi2008effect,yerushalmi2012atypical,yerushalmi2012students}.

However, such assumptions about advanced students' superior learning and self-monitoring skills have not been substantiated by research. Very little is known about whether a physics professor develops these skills in a continuous or discontinuous manner from the time they are introductory students.  There may be some discontinuous ``boosts'' in this process for many students, e.g., when they become involved in graduate research or when they ultimately independently start teaching and researching. There is also no research data on the fraction of students who have gone through the ``traditional'' undergraduate or graduate physics curriculum and have been unable to develop sufficient learning and self-monitoring skills, which are the hallmark of a physicist. 
%These issues are particularly important considering the diversity among STEM majors has increased significantly.

Moreover, investigations in which advanced physics students are asked to perform tasks related to simple introductory physics content do not properly assess their learning and self-monitoring skills \cite{chi1981categorization,larkin1980expert}.  Advanced students may possess a large amount of ``compiled knowledge'' about introductory physics and may not need to do much self-monitoring or learning while dealing with introductory problems. For example, when physics graduate students were asked to group together introductory physics problems based upon similarity of solution, their categorization was better than that of introductory physics students \cite{chi1981categorization}. While such tasks may be used to compare the grasp that introductory and advanced students have of introductory physics content, tasks involving introductory level content do not provide much insight into advanced physics students' learning and self-monitoring skills. 
The task of evaluating advanced physics students' learning and self-monitoring skills should involve advanced-level physics topics at the periphery of advanced students' own understanding. Also, while tracking the same student's learning and self-monitoring skills longitudinally is an extremely difficult task, taking snapshots of advanced students' learning and self-monitoring skills can be very valuable.

Earlier, Mason and Singh \cite{mason2010advanced} investigated the extent to which upper-level students in quantum mechanics learn from their mistakes. They administered four problems in the same semester twice, both on the midterm and final exams in an upper-level quantum mechanics course. The performance on the final exam shows that while some students performed equally well or improved compared to their performance on the midterm exam on a given question, a comparable number performed poorly both times or regressed (i.e., performed well on the midterm exam but performed poorly on the final exam). The wide distribution of students' performance on problems administered a second time points to the fact that many advanced students may not automatically exploit their mistakes as an opportunity for repairing, extending, and organizing their knowledge structure. Mason and Singh also conducted individual interviews with a subset of students to delve deeper into students' attitudes toward learning and the importance of organizing knowledge. In these individual interviews, they also found evidence that even in these advanced courses, many students do not automatically learn from their mistakes and they often resort to rote learning strategies for getting through the course. 
For example, they found that many students focused on selectively studying for the exams and did not necessarily look at the solutions provided by the instructor for the midterm exams to learn, partly because they did not expect those problems to be repeated on the final exam and/or found it painful to confront their mistakes. 

Similar to the benefits observed for introductory physics students, one instructional strategy that may help even advanced students is explicitly prompting them to learn from their mistakes by rewarding them for correcting the mistakes.
\cite{mason2008identifying,cohen2008identifying,mason2008identifying,singh2007physicsII,yerushalmi2007physics,yerushalmi2008effect,yerushalmi2012atypical,yerushalmi2012students}.
Indeed, giving incentives even to advanced students for learning from their mistakes, e.g., by explicitly rewarding them for correcting their mistakes can be an excellent learning opportunity for many students. Students may gain a new perspective on their own solutions, which may have mistakes, by asking themselves reflective questions while correctly solving the problems making use of the resources available to them. These issues are particularly important considering that the diversity in the prior preparation of  students at all levels has increased and many students need explicit guidance not only in the introductory courses, but also in the advanced courses.  

Here, we discuss a study spanning four years in which advanced undergraduate physics students taking a quantum mechanics course were given the same four problems in both the midterm exam and final exam similar to the Mason and Singh study but approximately half of the students were given incentives to correct their mistakes in the midterm exam and could get back up to 50\% of the points lost on each midterm exam problem.  The solutions to the midterm exam problems were provided to all students but those who corrected their mistakes were provided the solution after they submitted their corrections to the instructor.  The performance on the same problems on the final exam  suggests that students who were given incentives to correct their mistakes significantly outperformed those who were not given an incentive. 

\section{Methodology}

Our study took place over four years (the data were collected in four separate but identical courses) in the first semester of a two-semester upper-level undergraduate quantum mechanics course sequence taught by the same physics instructor at the University of Pittsburgh. This upper-level quantum mechanics course sequence is mandatory only for those students who want to obtain an honors degree in physics. It is often one of the last courses an undergraduate physics major takes.  Most students in this course are physics or engineering physics majors in their senior year (but some are in their junior year and there are also a few first year physics graduate students, who typically did not take a full year quantum mechanics sequence as undergraduate students). 
The four years in which the data were collected in the course were not consecutive years because typically a physics instructor at that university teaches a course for two consecutive years and then another instructor teaches it. Therefore, the data were collected from four classes that spanned a six year period. 

The classes were primarily taught in a traditional lecture format but the instructor had the students work on some preliminary tutorials that were being developed. Students were assigned weekly homework throughout the fifteen-week semester. In addition, there were two midterm exams and a final exam. The homework, midterm and final exams were the same in different years. The midterm exams covered only limited topics and the final exam was comprehensive. Students had instruction in all relevant concepts before the exams, and homework was assigned each week from the material covered in that week. Each week, the instructor held an optional class in which students could ask for help about any relevant material in addition to holding office hours. The first midterm exam took place approximately eight weeks after the semester started, and the second midterm exam took place four weeks after the first midterm examination. For our study, two problems were selected from each of the midterm exams and were given again verbatim on the final exam along with other problems not asked earlier.  The problems given twice are listed in Appendix \ref{IncentiveQuestions}.  

In the second and fourth year in which this study was conducted, the data were collected from classes in which students were asked to self-diagnose their mistakes on both their midterm exams in the course and could earn a maximum of 50\% of the points lost on each problem for submitting the corrected solution to each of the midterm exam problems. These classes formed the experimental or incentivized group. Including both years, there were 31 students in the incentivized group. There were no self-diagnosis activities in the first and third year in that students were not provided any grade incentive to diagnose their mistakes and submit the corrected solution to the midterm exam problems. The students in these two years formed the comparison group. There were 33 students in the comparison group including both years. 

All students were provided the solution to each midterm exam problems. Thus, students in both the experimental and comparison groups had the opportunity to learn from their mistakes before they encountered the four problems selected from the midterm exams on their final exam (as noted earlier, two problems were selected from each midterm exam).
However, for the experimental group, the midterm solutions were provided after students self-diagnosed their mistakes. Moreover, written feedback was provided to all students as needed in both the experimental and comparison groups after their midterm exam performance, indicating on the exams where mistakes were made. As noted, students in the incentivised group were asked to submit the corrected solution to each problem on the midterm exam on which they did not have a perfect score. They were given four days to diagnose and correct their mistakes and submit corrected solutions. They were directed to work on their own while correcting their mistakes, but were free to use any resources, homework, notes, and books to help them with this correction opportunity.  Of course, students in either the comparison group or incentivized group were free to use these resources to study at any time prior to the final in-class exam. 
 
 It is worth noting that these questions were in-class exam questions: short enough to be answered during the exam and similar to homework and quiz questions that students previously worked on. The corrected solution submitted by most students after the self-diagnosis was almost perfect so it was easy for the instructor to reward students with 50\% of the points lost. Our goal was to evaluate how students performed in subsequent problem solving based upon whether they diagnosed and corrected their mistakes on the midterm exam when provided with a grade incentive.

Three of the problems chosen (problem 1 which will also be called the expectation value problem for convenience, problem 2 or measurement problem and problem 3 or momentum problem in Appendix \ref{IncentiveQuestions}) were those that several students had difficulty with; a fourth problem (problem 4 or harmonic oscillator problem) which most students found straightforward on one of the two midterm exams was also chosen.  The most difficult of the four problems (based upon students' performance) was the momentum problem in Appendix \ref{IncentiveQuestions} that was also assigned as a homework problem before the midterm exam but was perceived by students to be more abstract in nature than the other problems.

\section{Rubrics and Scoring}

A scoring rubric, developed jointly with E. Yerushalmi and E. Cohen\cite{yerushalmi2012atypical,yerushalmi2012students} to assess how well the students in introductory physics courses diagnose their mistakes when explicitly prompted to do so, was adapted to score students' performance on each of the four quantum mechanics problems on both the midterm and final exams.  The scoring was checked independently by two scorers for approximately 25\% of the students and at least 95\% agreement was found on the scoring for each student on each problem in each attempt (on midterm and final exams).

The scoring rubric has two sections: one section scores students on their physics performance and the other section scores how well they presented their solution. The rubric for the presentation part was somewhat different from the corresponding part for introductory physics because quantum mechanics problems often asked for more abstract answers (e.g., proving that certain energy eigenstates are equally probable) as opposed to finding a numerical answer.  Therefore, some categories in the introductory physics rubric (e.g., writing down units) were omitted from the presentation part of the quantum mechanics rubric and other categories were adapted to reflect the nature of the quantum problems better (e.g., checking the answer was adapted to making a conceptual connection with the results).  

Although the grading rubric allows us to assign scores to each student for performance on physics and presentation parts separately, these two scores are highly correlated with the regression coefficient between the two scores being R=0.98. The reason for this high correlation is that students' presentation of the problem depended upon whether or not they understood the physical content. If a student did not know the relevant physical concepts, he/she could not set up an appropriate problem solving strategy to score well on the presentation part. We therefore only focus on students' physics scores on each of the four questions given on the midterm exam (called the pretest) and final exam (called the posttest). 

Appendix \ref{incentiveRubrics} demonstrates the scoring rubric for physics for all the four problems.  Below, we first describe the symbols used for scoring and then explain how a quantitative score is derived after the initial scoring is assigned symbolically for each sub-part of the rubric.  The symbol ``+'' (worth 1 point) is assigned if a student correctly completes a task as defined by the criterion for a given row.  The symbol ``-'' (worth 0 points) is assigned if the student either fails to do the given task or does it incorrectly.  If a student is judged to have gotten something partially correct, then the rater may assign a combination of pluses and minuses (++/-, +/-, +/--) to reflect this performance, with the understanding that such a combination represents an average score of pluses and minuses (e.g. ++/- translates to 2/3 of a point).  If the student's solution does not address a criterion then ``n/a'' (not applicable) is assigned and the criterion is not considered for grading purposes at all. For example, if the student does not invoke a principle,  the student will receive a ``-'' in the invoking appropriate concepts row but will receive ``n/a'' for applying it in the applying appropriate concept row because the student cannot be expected to apply a principle that he/she did not invoke. We note that ``using legitimate principles or concepts that are not appropriate in this problem" and ``using invalid principles and concepts (for instance, confusing a general state $\vert \psi \rangle$ with an eigenstate of an operator corresponding to an observable $\vert \psi_n\rangle$)" are included in invoking appropriate concepts category in Appendix \ref{incentiveRubrics}. However, a student will not lose points if he/she wrote legitimate principles or concepts but never used them to solve the problem. 

Although we only focus on students' physics scores, we note that the rubric for the presentation part included sub-components such as organization, plan, and evaluation. For example, for the momentum problem, the organization sub-component of the presentation part of the rubric included clear/appropriate knowns, for example, $e^{u } = \sum\limits_n \frac{1}{n!}{u}^n$. The plan sub-component of the rubric for momentum problem was divided into (1) appropriate target quantity chosen, (2)appropriate intermediate variables chosen, and (3)consistent plan. The evaluation sub-component of the rubric for the momentum problem included completing proof: $f(x+x_0) = e^{i\op{p}x_0 / \hbar}f(x)$ and making connection with results (momentum operator is generator of translation in space).

An overall or cumulative score is tabulated for each question for each student (see Appendices C and D for examples of scores using the rubric for student solutions from the comparison group and incentivized group on the pretest and posttest).  For the cumulative physics score for each student on a given problem on the midterm or final exams, the average of the scores for each subcategory (e.g., invoking appropriate and inappropriate physics concepts and applying concepts correctly or incorrectly) is used.

\section{Results}

All students who had less than perfect score on the midterm exams took advantage of the incentive to correct their mistakes for course credit. Our goal was to investigate correlation between the midterm exam score (which we call pretest score) and final exam score (which we call posttest score) on the four common problems for each group (comparison and incentivized groups). In particular, by comparing the performance of incentivized students with the comparison group, we examined the effect of giving grade incentives to correct one's mistakes on the pretests on subsequent performance on the posttest on the four problems repeated from the pretest. 

The comparison group and incentivized group had nearly identical average performance on the pretest as shown in 
Table \ref{tabIncentiveGains} (t-test comparing the pretest data of all students in the comparison and incentivized groups shows t=-0.037 and p=0.971).  A similar analysis comparing the posttest performance of the two groups shows that incentivized group performance is statistically significantly better than the comparison group (with t=3.265 and p= 0.002). In addition, we conducted a repeated measures ANOVA \cite{glass} and investigated the interaction between the groups and the growth from the pretest to the posttest scores and find a statistically significant difference (with F(1,62)=11.6, p=0.001, $\eta^2 =0.16$).  

\begin{table}[htbp!]
  \caption[Comparison Group vs Incentivized Group Gains]{Average pretest and posttest scores, gain and normalized gain ($g$) for students in the comparison and incentivized groups broken down into low, medium and high performance categories based on students' pretest score and for all students. While the pretest scores are comparable for the comparison group (number of students $N=33$) and incentivized group ($N=31$), the posttest scores are significantly higher for the incentivized group (reflected in gain and $g$). The numbers of students in the low, medium and high performance categories are 8, 14 and 11, respectively in the comparison group, and 7, 15 and 9, repsectively in the incentivized group. }
  \label{tabIncentiveGains}
  \centering
  \begin{tabularx}{\linewidth}{>{\itshape}l XXXX c XXXX}
  \toprule
 
  & \multicolumn{4}{c}{Comparison Group} & &\multicolumn{4}{c}{Incentivized Group} \\
  \hhline{~----~----}

  & Pre & Post & Gain&g& & Pre & Post & Gain&g \\
  \hhline{-----~----}  
   \multicolumn{1}{l}{\bfseries All}   &{\bfseries 67.9} &{\bfseries 71.5} &{\bfseries +3.6}& {\bfseries 0.112}  & &{\bfseries 67.6} &{\bfseries 88.4} &{\bfseries +20.8} & {\bfseries 0.642}\\
  \hhline{-----~----}   
   Low &34.6 &50.8 &+16.2& 0.248&   &30.8 &75.7 &+44.9&0.649 \\
 \hhline{-----~----}  
  Medium  &64.7 &66.3 &+1.6&0.045&    &67.3 &88.5 &+21.3&0.651 \\
 \hhline{-----~----}  
  High  &96.0 &93.0 &-2.9& -&   &96.9 &98.1 &+1.2&0.387 \\
  \hhline{-----~----} 
  
  \bottomrule
  \end{tabularx}
\end{table}

We also conducted a t-test to compare the differences between the pretest and postest scores for all students in each group. We find that for the comparison group, the difference between the pretest and postest scores is not significant (t=-0.590 and p=0.557) but for the inventivized group, the difference between the pretest and posttest scores is significant (t=-3.966 and $p < 0.001$). Moreover, the Cohen's d is 1.01 between the pretest and posttest performance of the incentivized group, which signifies a large effect size \cite{glass}. Cohen's d is given by \cite{glass} the difference between the means divided by the spooled standard deviation where the spooled standard deviation is defined as $\sigma_{spool}=\sqrt{(\sigma_{comparison}^2+\sigma_{incentivized}^2/2)}$ .

These results suggest that the comparison group students' average posttest performance on these problems is comparable to their performance on the pretest but for the students in the incentivized group, the average posttest performance on these problems is significantly better than their performance on the pretest. 
We also note that the fact that the average score on the posttest is comparable to the pretest in the comparison group suggests that the assumption that the upper-level physics students will automatically learn from their mistakes may not be valid. 

Examination of the performance of students in the comparison group shows that some students did well both times or improved in performance but others did poorly both times or deteriorated on the posttest.  We find that students struggled the most on the momentum problem both on the pretest and posttest. Moreover, examining the students who regressed from the pretest to posttest, we observe a pattern in which students who answered a question correctly on the pretest employed a different procedure for the same question on the posttest.  The procedure used was often a technique learned in the second half of the course and was not relevant to solving the problem.  This is suggestive that some students are applying memorized procedures, rather than trying to actually understand the problem they are solving. 

In the comparison group, in many of these cases in which students performed poorly, students wrote extensively on topics that were irrelevant to the question asked.  It is hard to imagine that students did not know that the things written by them were likely to be not relevant to the questions asked. It is possible that the students thought that if they wrote anything that they could remember about the topic (whether relevant or not) they may get some points on the exam for trying. Often, the irrelevant writings of a student on a particular question were different in the pretest and posttest. The poor performance of the students both times suggests that when the pretest was returned to them and the correct solutions were provided, they did not automatically use their mistakes as an opportunity for learning. A typical example of a student response for each question from the comparison group showing worse performance or comparable poor performance on the posttest as in the pretest (along with the scores tabulated using the rubric) is shown in Appendix C.

Figure \ref{figIncentivedGain} shows the average gain over the four problems (defined as the arithmetic difference between posttest and pretest averages for the questions; gain can therefore range from -100\% to +100\%) vs pretest score (which can range from zero to 100\%) for each student on the four questions repeated from the pretest to posttest in the comparison group and incentivized group.  Red-filled triangles are for each student from the comparison group and the blue-filled squares are used for each student in the incentivized group.  Students whose score improved are above the horizontal axis at zero gain; students whose performance deteriorated are below the horizontal axis.  The regions of possible gain are shaded according to posttest score performance categories: green for High posttest performance, yellow for Medium posttest performance and orange for Low posttest performance.  The performance categories are defined as follows: ``High,'' for scores from 85\% to 100\%; ``Medium'' for scores from 50\% to 85\%; and ``Low'' for scores from zero to 50\%.  The 50\% cutoff was chosen somewhat arbitrarily and the 85\% cutoff was chosen so that roughly one third of the students scored in the High performance category on the pretest.

Figure \ref{figIncentivedGain} and Table \ref{tabIncentiveGains} both show that students with poor performance on the midterm exam were likely to benefit from self-diagnosis activities in which they submitted the self-diagnosed corrected midterm exam solutions for 50\% of the points lost on each problem.  Therefore, the gap between the High and Low performers on the midterm exam was reduced for this incentivized group on the repeated problems on the final exam. On the other hand, for the comparison group in which students did not correct their mistakes but were given the correct solutions to all of the midterm exam problems, the gap remained, i.e., scores did not substantially improve for low performers, and remained diverse.

\begin{figure}[htb!]

\begin{center}
\includegraphics[width=\linewidth]{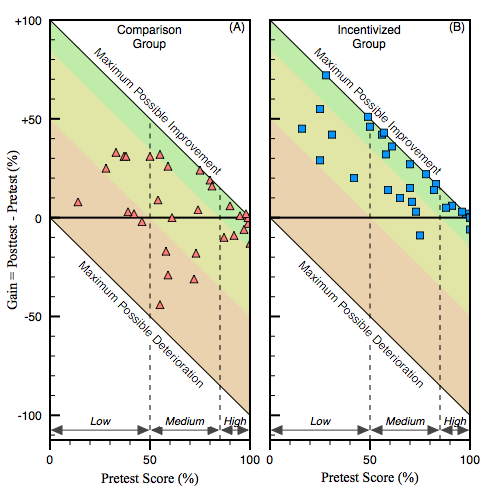}
  \caption[Average Gain vs Pretest Performance for Incentivized and Comparison Groups]{Average gain (defined as the difference between posttest and pretest score) vs pretest score for each student on the four questions repeated from the pretest to posttest in the comparison group (A) and incentivized group (B).  Red-filled triangles are for each student from the comparison group and the blue-filled squares are for each student in the incentivized group in which students received an explicit grade incentive to correct their own mistakes in pretest before answering the same questions on the posttest.  Students whose scores improved are above the horizontal axis; students whose performance deteriorated are below the horizontal axis.}
  \label{figIncentivedGain}
\end{center}
\end{figure}

As shown in both Figure \ref{figIncentivedGain} and Table \ref{tabIncentiveGains}, the data were analyzed by breaking the students into three groups based on their pretest performance. The initially high-performing students from both the comparison and incentivized groups (scoring 85\% and higher on the pretest) generally performed very well on the posttest regardless of the intervention (see Figure \ref{figIncentivedGain}).  Most of these students who start in the High pretest category stay in that category.  Students who initially performed at a Medium level on the pretest (scoring between 50\% and 85\%) in the incentivized group perform better on the posttest than the corresponding students who were in the comparison group.  In the comparison group, students in the Medium performance category on the pretest were as likely to improve on the posttest (above the horizontal axis in Figure \ref{tabIncentiveGains}A) as they were to deteriorate (below the horizontal axis).  In contrast, in the incentivized group, almost all of the students in the Medium category on the pretest improved on the posttest (see Figure \ref{tabIncentiveGains}B). 

 Furthermore, about half of these students in the incentivized group improved as much as possible on the posttest, saturating the boundary for maximal improvement (see Figure \ref{tabIncentiveGains}B). Among the initially Low performing students (pretest scores less than 50\%), many students in both comparison and incentivized groups improved on the posttest.  However, the degree to which these struggling students performed on the posttest is highly dependent on whether or not they received a grade incentive to improve.  The students in the Low category on the pretest in the incentivized and comparison groups had an average gain of 44.9\% and 16.2\%, respectively (Table \ref{tabIncentiveGains}). In summary, the gains are much larger for the incentivized group, bringing the average of the Low category to the level of the Medium category, and the Medium category to the High category (see Figure \ref{figIncentivedGain} and Table \ref{tabIncentiveGains}). A typical example of a student response for each question showing improvements in the incentivized group (along with the scores tabulated using the rubric) is shown in Appendix D.

The data shown in Figure 1 and Table I are shown in two other forms in Figure 2 because it may be instructive to view the data in another form. In particular, the average pretest vs. posttest score for each student on the four questions repeated from the pretest to posttest is plotted for the comparison group and incentivized group.  Red-filled triangles are used for each student from the comparison group and the blue-filled squares are for each student in the incentivized group.  In addition, the histograms along each axis show the distribution of the pretest and posttest scores for the comparison group and the incentived group. A closer look at the histograms suggests that pre-intervention scores are similar for the comparison and incentivized group.  However, the incentivized group posttest distribution has shifted out of the lower region, and is peaked near the High performance cutoff.

 Table \ref{tabIncentiveGains} also includes the average normalized gain~\cite{hake}, $g$, for each group for each performance category. The normalized gain is defined as the posttest percent minus the pretest percent divided by (100-pretest percent). The normalized gain scales the gain based upon the pretest scores (e.g., $g=0$ implies that there is no gain at all from the pretest to posttest, $g=0.5$ implies that the gain is $50\%$ of the maximum possible that could have been gained and $g=1$ implies that the gain is the maximum amount possible. Table \ref{tabIncentiveGains} shows that the normalized gain $g$ for each pretest performance category is larger for the incentivized group than the comparison group (negative gains are not converted to the normalized gain \cite{hake}). For incentivized group, the normalized gain is large for the low and medium pretest performance categories.

\begin{figure}[htb]
  \caption[pretest vs. posttest]{Average pretest vs. posttest score for each student on the four questions repeated from the pretest to posttest in the comparison group and incentivized group.  Red-filled triangles are used for each student from the comparison group and the blue-filled squares are for each student in the incentivized group.  The histograms along each axis show the distribution of the pretest and posttest scores for the comparison group and the incentived group. A closer look at the histograms suggests that pre-intervention scores are similar for the comparison and incentivized group.  However, the  incentivized group posttest distribution has shifted out of the lower region, and is peaked near the ``High'' performance cutoff.}

\begin{center}
\includegraphics[width=\linewidth]{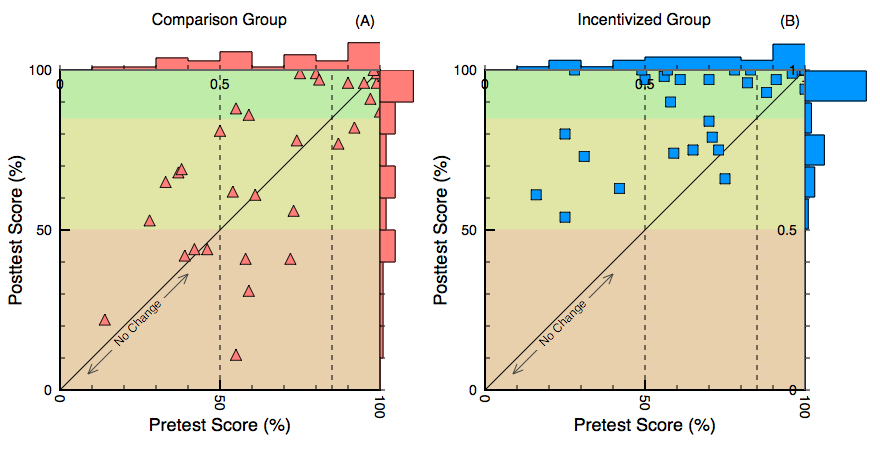}
\end{center}
\end{figure}

While these results are encouraging, caution is urged in interpreting improvement.  Our findings support the claim that students improve on problems administered a second time when expert-like behavior (self-diagnosing and correcting their mistakes) is explicitly incentivized.  It does not necessarily follow that students have become adept at self-monitoring skills from just two such interventions in the two midterm exams in quantum mechanics. In particular, we compared the performance of students in the incentivized and comparison groups on four other problems on the same final exam for which incentivized group students did not diagnose their mistakes. We find that while the incentivized group scored higher than the comparison group, the results are not statistically significant.

\section{Discussion}A common hypothesis is that advanced students will benefit from solutions to exams provided after the exams because they would want to learn from their mistakes.  However, we find that many upper-level students who do not receive an explicit grade incentive to learn from their mistakes, in fact, do not learn from their mistakes.  More encouragingly, students who are given an incentive (earn back some points for correcting their mistakes), typically perform substantially better on problems repeated a second time from midterm to final exam.

Prior research demonstrates that many introductory students do not automatically learn from their mistakes without explicit intervention \cite{cohen2008identifying,mason2008identifying,singh2007physicsII,yerushalmi2007physics,yerushalmi2008effect,yerushalmi2012atypical,yerushalmi2012students}. Moreover, many introductory physics students are ``captive audiences'' -- they may not buy into the goals of the course and their main goal becomes getting a good grade even if their learning is superficial \cite{elby2001helping}. Research suggests that the introductory physics students can benefit from explicit guidance and feedback in developing problem solving and learning skills and alignment of course goals with assessment methods \cite{elby2001helping,eylon1984effects,meltzer2005relation,ozimek2005retention,van1991overview,cohen2008identifying,mason2008identifying,singh2007physicsII,yerushalmi2007physics,yerushalmi2008effect,brewe2008modeling,hestenes1987toward,yerushalmi2012atypical,yerushalmi2012students}.
However, it is commonly assumed that the learning skills of students in advanced physics courses are superior to those of students in introductory courses so they will monitor their learning and learn from their mistakes.  For example, instructors often believe that physics seniors in an upper-level quantum mechanics course can monitor their own learning and they will automatically take the time out to learn from their mistakes. One reason offered is that advanced physics majors have chosen this major and are therefore eager to learn the material. They will make every effort to repair, organize and extend their knowledge because they are intrinsically motivated to learn and are not grade driven. 

Contrary to these beliefs, our earlier investigation \cite{mason2010advanced} found that the performance of advanced students in the upper-level quantum mechanics sequence did not automatically improve on identical questions given on a midterm exam and on the final exam.  The students were provided the correct solutions and their own graded exams. Even then, there was an apparent lack of reflective practice by supposedly mature students and many students did not take the opportunity to repair and organize their knowledge structure. In individual interviews, we probed students' attitudes and approaches towards problem solving and learning, and also asked them to solve the same problems again. The statistical results were consistent with some students' ``self-described'' approaches towards problem-solving and learning. In the interviews, we find evidence that even in these advanced courses there are students who do not use their mistakes as an opportunity for learning and for building a robust knowledge structure; they resort to rote learning strategies for getting through the course. For example, one interviewed student alluded to the fact that he always looked at the correct homework solutions provided but did not always look up the correct midterm exam solutions partly because he did not expect those questions to be repeated on the final exam. This tendency to "study" the problems that may show up on the exam without making an effort to build a good knowledge structure is typically not expected of physics students in advanced courses. 

Individual discussions with some physics faculty suggests that sometimes their incorrect inferences about advanced physics students' learning and self-monitoring skills are based on the fact that they feel that all physics majors are like them. They may not appreciate the large diversity in the population of physics majors today and may not realize that those who become college physics faculty consist of a very select group of undergraduate physics majors. While longitudinal research is needed to investigate the differences between those advanced students who pursue graduate study and eventually become physics faculty and those who do not, it is possible that those students aspiring to be physics faculty are intrinsically motivated to learn and make more effort to learn from their mistakes on their own. 

Our study suggests that similar to introductory students, advanced physics students who do not automatically use their mistakes as a learning opportunity may benefit from explicit scaffolding support and guidance to help them learn from their mistakes. Students will automatically use problem solving as an opportunity for reflecting and learning if they are intrinsically motivated to learn the content and to extend and organize their knowledge \cite{ames1992classrooms,dweck1988social,pintrich1999role,stipek2002motivation}. However, advanced students who are not intrinsically motivated may need extrinsic motivation, e.g., explicit reward to help them learn from their mistakes. The strategy discussed here is explicitly asking advanced students in upper-level quantum mechanics course to correct their mistakes in midterm exams similar to the strategies that have been used successfully for introductory courses \cite{cohen2008identifying,mason2008identifying,singh2007physicsII,yerushalmi2007physics,yerushalmi2008effect,yerushalmi2012atypical,yerushalmi2012students}.
Our research suggests that even many upper-level students may be more motivated to engage with instructional material in a more meaningful way if they are provided a grade incentive to correct their mistakes. 
Considering the relative ease with which instructors in physics courses at all levels can implement the intervention in which students are given grade incentives to correct and learn from their mistakes, instructors at all levels should consider giving students this opportunity to learn. Asking students to correct their mistakes in many courses may also help students understand the importance of learning from mistakes and the role of appropriate productive struggle in learning physics. 

Many research-based instructional approaches require training for instructors, instructor preparation time, and materials which can add to costs associated with teaching.  However, the self-diagnosis and correction of mistakes discussed in this paper does not require exceptional effort on the part of instructors, nor does it require class time.  Like many effective research-based methods, the student is the active agent.  It has been found that for introductory students, the corrections themselves take some time to evaluate, incurring some ongoing instructor labor\cite{yerushalmi2012atypical,yerushalmi2012students}.  However, for upper-level students in quantum mechanics, we find that students seize the opportunity and produce expert-like solutions when asked to diagnosed their mistakes and submit corrected solutions for course credit.   Therefore, the additional instructor labor is minimal, due to the high quality of the solutions that students typically produce after self-diagnosis of mistakes.  In any case, the additional labor on the part of the instructor if students are asked to diagnose their own mistakes for a grade incentive is no more than an additional round of exam evaluations. Moreover, the self-diagnosis of mistakes discussed in this research study can be used along with other research-based curricula and pedagogies. 

\section{Conclusion}
  
Explicitly providing grade incentives to correct mistakes on the midterm exam positively affected the final exam performance of students with a diverse spectrum of prior performance on the same problems and the students who were given incentives to correct their mistakes significantly outperformed those who were not given an incentive.  In other words, the performance of students in the group in which no incentives were provided shows that while some advanced students performed equally well or improved compared to their performance on the midterm exam on the questions administered a second time, a comparable number of students obtained lower scores on the final exam than on the midterm exam. The wide distribution of students' performance on problems administered a second time in this case suggests that many advanced students do not automatically exploit their mistakes as an opportunity for learning. 
An explicit incentive to correct their mistakes can be an effective formative assessment tool \cite{blackwil,brown}. 
If this type of easy-to-implement intervention is implemented routinely in all physics courses, students are likely to use their mistakes as a learning opportunity and may even develop better self-monitoring skills over time. 
We note that we are advocating for instructors to give students incentives to correct their mistakes in all STEM courses; we are not suggesting that they repeat the same questions on a final exam as was necessary for the purposes of this research.

\begin{acknowledgments}
We thank the National Science Foundation for Awards No. PHY-1505460 and No.PHY-1202909 and F. Reif, J. Levy and R. P. Devaty for extremely useful discussions.
\end{acknowledgments}

\bibliographystyle{aipproc}

\clearpage
\appendix
\section{Questions}\label{IncentiveQuestions}
\newcommand{\surveyItem}[1]{\item \begin{minipage}[t]{\linewidth} #1 \end{minipage}}

The following problems were given both on the midterm and final exams. Students were given an additional sheet on which useful information was provided. For example, they were given the explicit form of $\psi_n(x)$ and $E_n$ (the $n^{th}$ energy eigenfunction and the $n^{th}$ energy eigenvalue, respectively) for a one-dimensional (1-D) infinite square well. For the 1-D Harmonic Oscillator, they were given the energies $E_n$ in terms of quantum number $n$, how the lowering and raising operators, $a_{-}$ and $a_{+}$, relate to the position and momentum operators, \op{x} and \op{p}, the commutation relation between the raising and lowering operators, $[a_{-}, a_{+}]$, and how the raising and lowering operators,  $a_{+}$ and $a_{-}$, acting on the $n^{th}$ energy eigenstate of a one-dimensional harmonic oscillator changes that state etc.  

{\noindent
1) The eigenvalue equation for an operator \op{Q} is given by  $\op{Q}\ket{\psi_i} = \lambda_i \ket{\psi_i} $, with $i = 1...N $.  Find an expression for $\langle \psi \vert \hat Q \vert \psi \rangle$, where $\vert \psi \rangle$
 is a general state, in terms of $\langle \psi_i \vert \psi \rangle$.

{\noindent
2) For an electron in a one-dimensional infinite square well with well boundaries at $x=0$ and $x=a$, measurement of position yields the value $x=a/2$.  Write down the wave function immediately after the position measurement and without normalizing it show that if energy is measured immediately after the position measurement, it is equally probable to find the electron in any odd-energy stationary state.
}

{\noindent
3) Write an expression to show that the momentum operator $\op{P}$ is the generator of translation in space.  Then prove the relation.  (Simply writing the expression is not sufficient... you need to prove it.)
}

{\noindent
4) Find the expectation value of potential energy in the $n^{th}$ energy eigenstate of a one dimensional Harmonic Oscillator using the ladder operator method.
}
\clearpage
\section{Rubrics}\label{incentiveRubrics}
Summaries of the rubrics used to evaluate student performance on the physics part on all four problems are shown here.  As noted in the text, each problem receives an overall score, which is the average of the “Invoking” and “Applying” scores.  Each of these general criteria scores is in turn the average of the specific criteria scores, which can take on the values 0, 1, $1/3$, $1/2$, $2/3$, or N/A.  “N/A” arises when certain criteria are not present to include in determining a score.  Inter-rater reliability was tested for roughly 25\% of the dataset and found to be better than 95\%.
\newlength{\rubricOverall}
\setlength{\rubricOverall}{0.12\linewidth}
\newlength{\rubricGeneral}
\setlength{\rubricGeneral}{0.13\linewidth}

\begin{table*}[tbh!]
  
  \setlength{\tabcolsep}{.01\linewidth}

  \abovedisplayskip=6pt
  \belowdisplayskip=6pt
  \abovedisplayshortskip=0pt
  \belowdisplayshortskip=3pt

  \caption[Summary of the Rubric Used for Problem 1 (expectation value problem).]{Summary of the rubric used for problem 1 (expectation value problem).  In this problem, students are asked to write the expectation value of an observable $Q$ in terms of eigenstates and eigenvalues of the corresponding operator.}
  \label{tabIncentiveRubric1}

  \centering
  \begin{tabularx}{\linewidth}{|m{\rubricOverall}|m{\rubricGeneral}|X|}
  \cline{2-3}
  \multicolumn{1}{l|} {Problem 1} & {General Criteria} & {Specific Criteria}\\
  \hline

  \multirow{8}{\rubricOverall}{Overall Score} & 

  \multirow{5}{\rubricGeneral}{Invoking appropriate concepts} & 
  {Spectral decomposition expressing identity operator in terms of a complete set of eigenstates $\ket{\psi_n}$:}
  \[  \op{I} = \sum \ket{\psi_n}\bra{\psi_n} \]
 {Or expressing general state in terms of the eigenstates of $\op{Q}$:}
 \[\ket{\psi} = \sum c_n \ket{\psi_n} \text{, where }c_n = \langle \psi_n | \psi \rangle \]\\
 \cline{3-3}
 & & Make use of $\op{Q}\ket{\psi_n} = \lambda_n\ket{\psi_n}$ \\
 \cline{3-3}
 & & ${\langle \psi_n | \psi \rangle}^{*} = \langle \psi | \psi_n \rangle$ \\
 \cline{3-3}
 & & Using legitimate principles or concepts that are not appropriate in this problem.\\
 \cline{3-3}
 & & Using invalid principles or concepts (for instance, confusing a general state ￼ $\ket{\psi}$ with
an eigenstate $\ket{\psi_n}$).\\
\cline{2-3}

& \multirow{3}{\rubricGeneral}{Applying \newline concepts} & 

Inserting spectral decomposition into the expression for expectation value\\ \cline{3-3}

& & Eigenvalue evaluated and treated as number.\\ \cline{3-3}

& & Probability expressed in terms of ${\langle \psi_n | \psi \rangle}^{*}$ and ${\langle \psi_n | \psi \rangle}$\\
\hline
\end{tabularx}
\end{table*}

\begin{table*}[tbhp]
 
  \setlength{\tabcolsep}{.01\linewidth}
 
  \abovedisplayskip=6pt
  \belowdisplayskip=6pt
  \abovedisplayshortskip=0pt
  \belowdisplayshortskip=3pt

  \caption[Summary of the Rubric Used for Problem 2.]{Summary of the rubric used to solve problem 2 (measurement problem).  In this problem, a particle in a one dimensional infinite square well is first measured to be at the center of the well.  Students are asked to show that a subsequent energy measurement can yield any odd-integer energy state with equal probability.}
  \label{tabIncentiveRubric2}

  \resizebox{0.9\linewidth}{0.34\paperheight}{
  \begin{tabularx}{\linewidth}{|m{\rubricOverall}|m{\rubricGeneral}|X|}
  \cline{2-3}
  \multicolumn{1}{l|} {Problem 2} & {General Criteria} & {Specific Criteria}\\
  \hline

  \multirow{14}{\rubricOverall}{Overall Score}

    & \multirow{6}{\rubricGeneral}{Invoking appropriate concepts}

      & Measurement of position yields a Dirac delta function for the wave function immediately after the position measurement. \\ 
       \cline{3-3}

  & & Expand the wavefunction in terms of energy eigenfunctions: \(\psi(x) = \sum c_n \psi_n(x)\) \\
      \cline{3-3}

  & & Express probability amplitude for energy measurement as:
  \(c_n = \int_{0}^{a} \psi_n^{*}(x)\psi(x) \dd x \) \\ 
      \cline{3-3}

  & & Express the probability of measuring a given energy \(E_n\) as \(\abs{c_n}^2\) \\ 
      \cline{3-3}

  & & Using legitimate principles or concepts that are not appropriate in this problem, e.g. invoking expectation values. \\ 
      \cline{3-3}

  & & Using invalid principles or concepts (for instance, confusing position eigenstates with energy eigenstates).\\ 
    \cline{2-3}

  & \multirow{8}{\rubricGeneral}{Applying concepts} 

    & Applying delta function definition correctly to write the wavefunction after the position measurement:\\

  & & $\psi(x) = A \delta(x-\frac{a}{2})$\\ 
      \cline{3-3}
  & & Using provided stationary states for infinite square well:\\
  & &  $\psi_n(x) = 
                  \begin{cases}
                    \sqrt{\frac{2}{a}}\sin(\frac{n \pi x}{a}) & \text{for } 0 \leq x \leq a\\  
                    0 & \text{otherwise}
                  \end{cases}$
        \\
      \cline{3-3}
  & & Dirac delta function identity applied appropriately to calculate probability amplitude for energy measurement:\\
  & & $c_n = A \sqrt{\frac{2}{a}} \int_0^a \sin(\frac{n\pi x}{a})\delta(x-\frac{a}{2}) \dd x$  \\
      \cline{3-3}

  & & Find the probability for measuring energy $E_n$,\\
  & & 
     $\abs{c_n}^2 = \abs{A}^2 \frac{2}{a} \sin^2(\frac{n\pi}{2})$
     \\

  & & $
     \hphantom{\abs{c_n}^2 } = \abs{A}^2\frac{2}{a} 
     \begin{cases}
     1 & \text{for $n$ odd}\\
     0 & \text{for $n$ even}
     \end{cases}
    % \end{flalign*}
    $
    \\
\hline
\end{tabularx}
}
\end{table*}

\begin{table*}[tbhp]
  
  \caption[Summary of the Rubric Used for Problem 3.]{Summary of the rubric used to solve problem 3 (momentum problem). In this problem, students prove that momentum operator is the generator of translation in space.}\label{tabIncentiveRubric3}

  \begin{tabularx}{\linewidth}{|m{\rubricOverall}|m{\rubricGeneral}|X|}
  \cline{2-3}
  \multicolumn{1}{l|} {Problem 3} & {General Criteria} & {Specific Criteria}\\
  \hline

  \multirow{9}{\rubricOverall}{Overall Score}
  & \multirow{6}{\rubricGeneral}{Invoking appropriate concepts}
    & Taylor expansion definition \\
  & & $f(x+x_0) = \sum\limits_{n=0}^\infty \frac{1}{n!}{x_0}^n\dv{x}^n f(x)$\\
      \cline{3-3}
  & & Momentum operator in position space in one dimension is $\op{p} = \frac{\hbar}{i}\dv{x}$ \\
      \cline{3-3}
  & & Expansion of exponential: $e^u = \sum\limits_{n=0}^\infty \frac{1}{n!} u^n$ \\
      \cline{3-3}
  & & Using legitimate principles or concepts that are not appropriate in this problem.\\
      \cline{3-3}
  & & Using invalid principles or concepts (for instance, confusing position space with momentum space).\\
      \cline{2-3}
  & \multirow{3}{\rubricGeneral}{Applying concepts}
    & Partial derivative in terms of momentum operator: $\pdv{x} = i \op{p} / \hbar$ \\
      \cline{3-3}
  & & $e^{i\op{p}x_0 / \hbar} = \sum\limits_n \frac{1}{n!}{x_0}^n(\frac{i\op{p}}{\hbar})^n$ \\
      \cline{3-3}
  & & Taylor expansion performed correctly to obtain:
      $f(x+x_0) = e^{i\op{p}x_0 / \hbar}f(x)$\\
  \hline
\end{tabularx}
\end{table*}

\begin{table*}[tbhp]
  % \singlespace
  \caption[Summary of the Rubric Used for Problem 4.]{The summary of the rubric used to solve problem 4 (harmonic oscillator problem).  In this problem, students are asked to find the expectation value of the potential energy for a one-dimensional harmonic oscillator when the system is in the $n^{th}$ energy eigenstate.}\label{tabIncentiveRubric4}

  \begin{tabularx}{\linewidth}{|m{\rubricOverall}|m{\rubricGeneral}|X|}
  \cline{2-3}
  \multicolumn{1}{l|} {Problem 4} & {General Criteria} & {Specific Criteria}\\
  \hline

  \multirow{9}{\rubricOverall}{Overall Score}
  & \multirow{6}{\rubricGeneral}{Invoking appropriate concepts}
    &$V = \frac{1}{2} m \omega^2 x^2 = \frac{1}{2} k x^2$ \\
  & & - or -\\
  & & $H = \frac{p^2}{2m} + \frac{1}{2}m \omega^2 x^2$\\
      \cline{3-3}
  & & Express expectation value of $f(x)$ as: \\
  & & ${\langle f(x) \rangle} = \int\limits_{-\infty}^\infty \psi^{*}(x) f(x) \psi(x) \dd x$\\
      \cline{3-3}
  & & Describe \op{x} (or \op{H}) in terms of raising and lowering operators $a_{+}$ and $a_{-}$ (as given in the formula sheet provided to students) \\
      \cline{3-3}
  & & Use orthogonality principle:$\int\limits_{-\infty}^\infty \psi_m^{*}(x) \psi_n(x) \dd x = \delta_{mn}$ \\
  & & Using legitimate principles or concepts that are not appropriate in this problem, e.g. $\psi_n$ in terms of $\psi_0$.\\
      \cline{3-3}
  & & Using invalid principles or concepts (for instance, an incorrect definition of expectation value).\\
      \cline{2-3}
  & \multirow{3}{\rubricGeneral}{Applying concepts}
    & Proper expansion of expressions, e.g. order of
ladder operators in cross terms $a_{+}a_{−}$ and $a_{-}a_{+}$ is correctly accounted for.\\
      \cline{3-3}
  & & Apply the operators correctly to obtain the correct states and coefficients. For instance, \\
  & & $a_{+} \ket{\psi_n} = \sqrt{n+1}\ket{\psi_{n+1}}$\\
      \cline{3-3}
  & & Apply the orthogonality relation:  $\int\limits_{-\infty}^{\infty}\psi_n^{*}(x)\psi_{n\pm2} (x)\dd{x} = 0$, $\int\limits_{-\infty}^{\infty}\psi_n^{*}(x)\psi_{n} (x) \dd{x} = 1$\\
  \hline
\end{tabularx}
\end{table*}

\clearpage

\section{Example Responses from the Comparison Group}
Below, we provide typical sample student responses from the comparison groups to show how students deteriorated from the pretest to posttest on the same problem.  Each example includes both pretest and posttest for a given student for a particular problem.

\begin{figure*}[tbhp!]
  \centering
  \includegraphics[width=\linewidth]{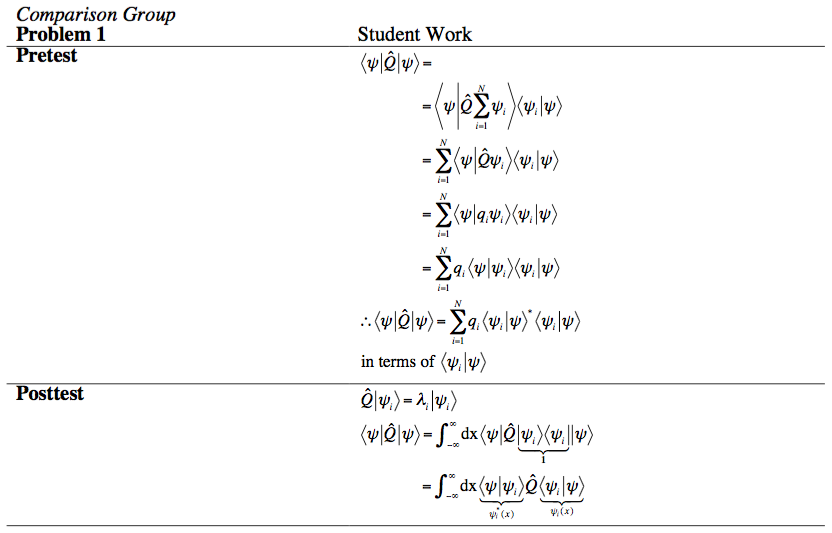}
  \caption{An example of a pretest and posttest solution pair for a student from the comparison group which demonstrates a typical deterioration on the expectation value problem. The pretest and posttest scores are 100\% and 50\% for invoking, 89\% and 33\% for applying, and 95\% and 42\% for the overall scores, respectively.  While the pretest solution is almost perfect, the posttest solution shows student difficulties including confusion between the discrete spectrum of the eigenstates of \op{Q} and the continuous spectrum of position eigenstates and difficulty with the treatment of the operator \op{Q}.}\label{tabIncentiveExampleC1}
\end{figure*}

\begin{figure*}[tbhp!]
  \centering
  \includegraphics[width=\linewidth]{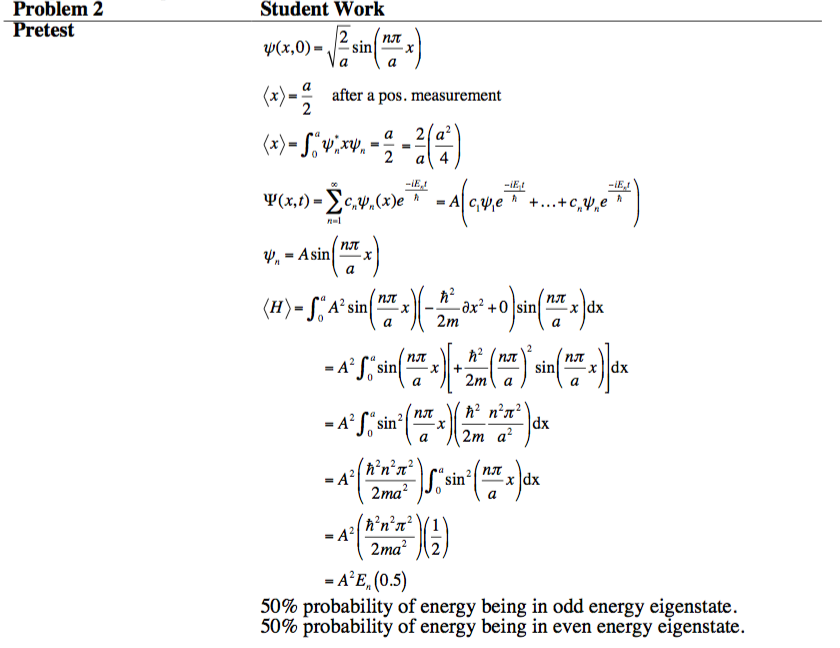}
  \includegraphics[width=\linewidth]{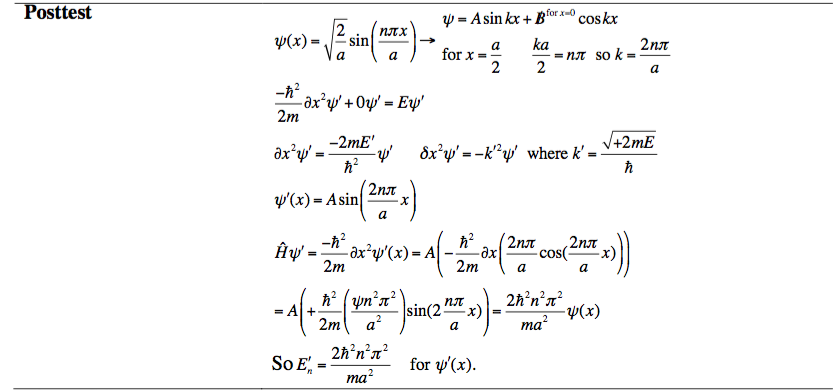}
  \caption{An example of a pretest and posttest solution pair for a student from the comparison group for the measurement problem. The pretest and posttest scores are 36\% and 8\% for invoking, 13\% and 17\% for applying, and 24\% and 13\% for the overall scores, respectively. The student struggles with this problem in the pretest. In the posttest, no improvement is demonstrated.}\label{tabIncentiveExampleC2}
\end{figure*}

\begin{figure*}[tbhp!]
  \centering
  \includegraphics[width=\linewidth]{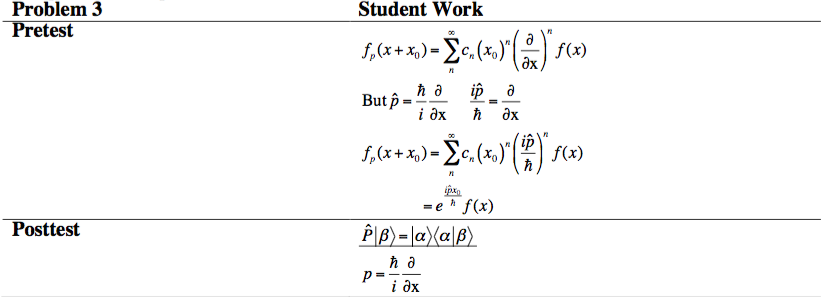}
  \caption{An example of a pretest and posttest solution pair for a student from the comparison group which demonstrates a typical deterioration for the momentum problem.  The pretest and posttest scores are 89\% and 0\% for invoking, 100\% and 0\% for applying, and 95\% and 0\% for the overall scores, respectively. The pretest solution is nearly perfect, but in the posttest, no relevant knowledge is displayed.}\label{tabIncentiveExampleC3}
\end{figure*}

\begin{figure*}[tbhp!]
  \centering
  \includegraphics[width=\linewidth]{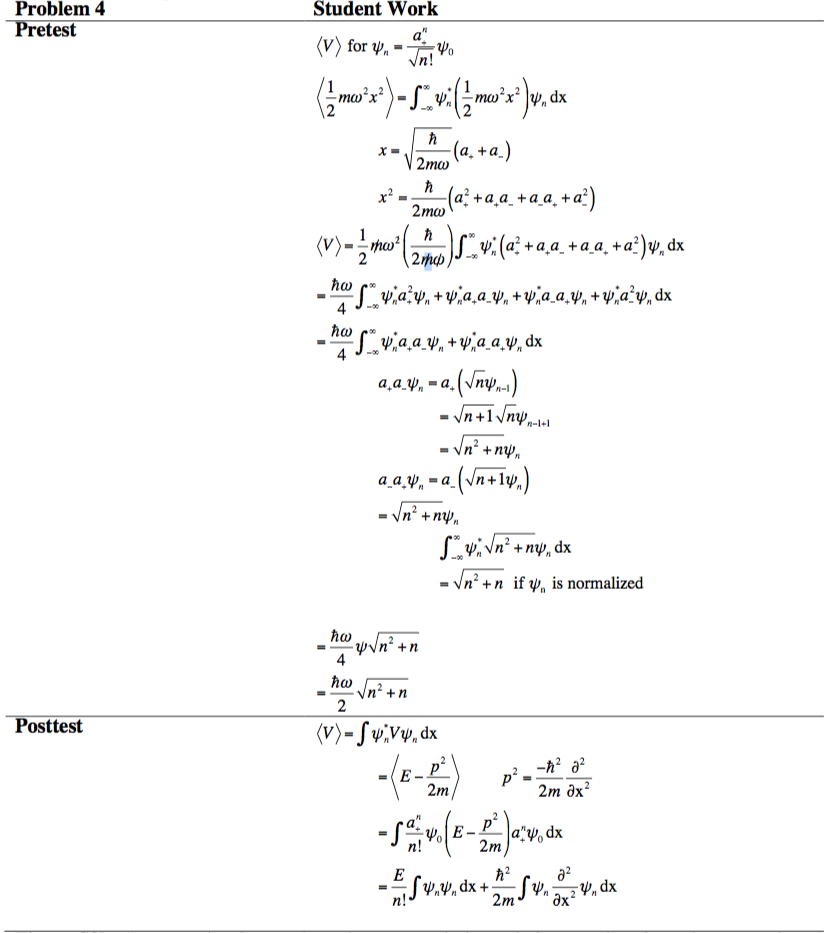}
  \caption{An example of a pretest and posttest solution pair for a student from the comparison group which demonstrates a typical deterioration for the harmonic oscillator problem.  The pretest and posttest scores are 100\% and 40\% for invoking, 83\% and 0\% for applying, and 92\% and 20\% for the overall scores, respectively. The student's pretest performance is good, except for some minor errors with the coefficients when dealing with the ladder operators but the posttest solution demonstrates that this proficiency has deteriorated.}\label{tabIncentiveExampleC4}
\end{figure*}

\clearpage
\section{Example Responses from the Incentivized Group}
Below, we provide typical sample student responses from the incentivized group to show how students improved from pretest to posttest. Each example includes both pretest and posttest for a given student for a particular problem.

\begin{figure*}[tbh!]
  \centering
  \includegraphics[width=\linewidth]{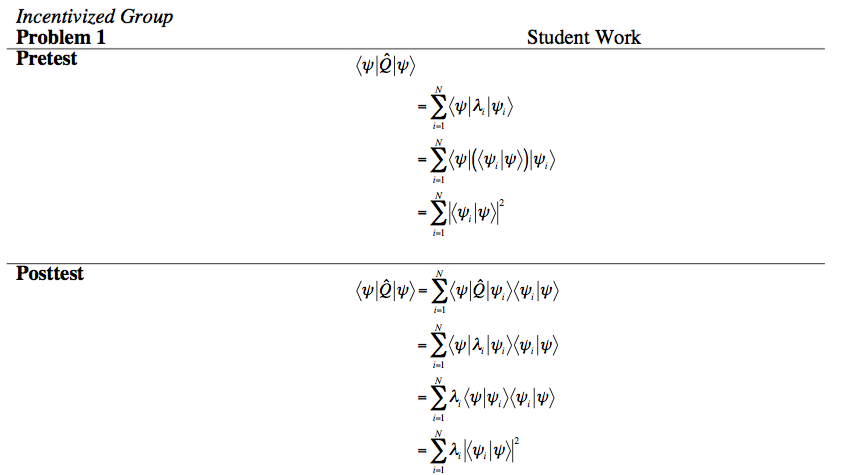}
  \caption{An example of a pretest and posttest solution pair for a student from the incentivized group which demonstrates improvement in student understanding for the expectation value problem. The pretest and posttest scores are 54\% and 100\% for invoking, 33\% and 100\% for applying, and 44\% and 100\% for the overall scores, respectively.  While the pretest solution shows difficulty with the expansion of a general state in terms of the eigenstates of the operator \op{Q} and confusion between the eigenvalue of \op{Q}  and the probability amplitude for measuring $Q$, the posttest shows excellent use of Dirac notation to solve the problem.}\label{tabIncentiveExampleI1}
\end{figure*}

\begin{figure*}[tbhp!]
  \centering
  \includegraphics[width=\linewidth]{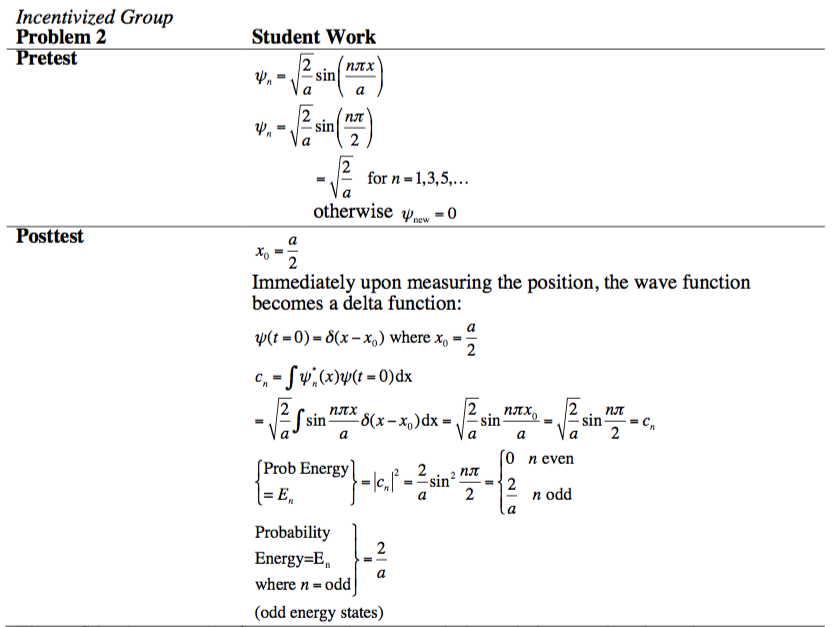}
  \caption{An example of a pretest and posttest solution pair for a student from the incentivized group which demonstrates improvement in student understanding for the measurement problem. The pretest and posttest scores are 0\% and 100\% for invoking, 25\% and 100\% for applying, and 13\% and 100\% for the overall scores, respectively. The pretest solution shows that the student incorrectly inserts $x=a/2$  in the expression for the energy eigenfunction instead of calculating the probability of measuring energy.  The posttest solution is essentially perfect.}\label{tabIncentiveExampleI2}
\end{figure*}

\begin{figure*}[tbhp!]
  \centering
  \includegraphics[width=\linewidth]{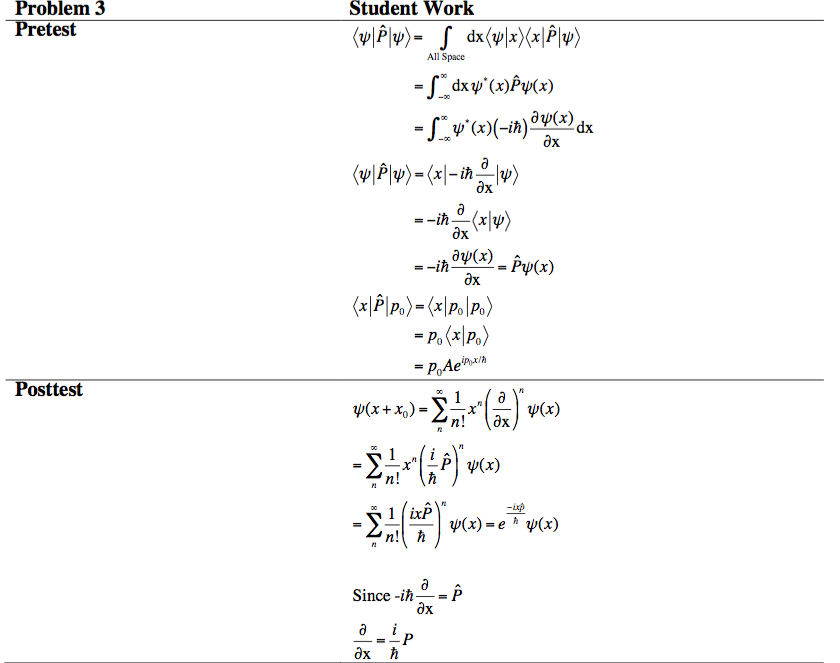}
  \caption{An example of a pretest and posttest solution pair for a student from the incentivized group which demonstrates improvement in student understanding for the momentum problem.  The pretest and posttest scores are 20\% and 100\% for invoking, 0\% and 89\% for applying, and 10\% and 95\% for the overall scores, respectively. The pretest shows exploratory derivations focused on expectation value of momentum that has nothing to do with the correct solution.  The posttest solution is a succinct derivation of the desired result.}\label{tabIncentiveExampleI3}
\end{figure*}

\begin{figure*}[tbhp!]
  \noindent

  \begin{tabular}{>{\bfseries}l>{\bfseries}l}
  Problem 4 & Student Work\\
  \toprule
  Pretest & Posttest \\
  \includegraphics[scale=0.5]{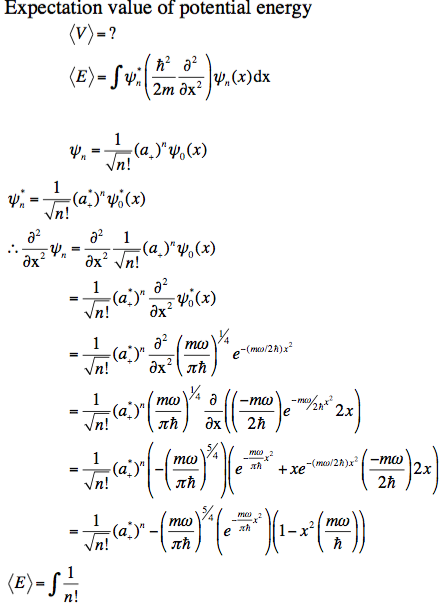} & \includegraphics[scale=0.5]{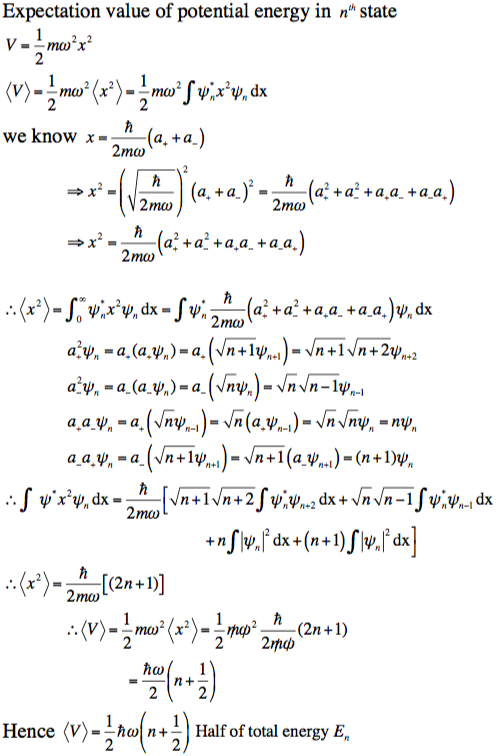} \\
  \end{tabular}

  \caption{An example of a pretest and posttest solution pair for a student from the incentivized group which demonstrates improvement in student understanding for the harmonic oscillator problem. The pretest and posttest scores are 33\% and 100\% for invoking, 0\% and 100\% for applying, and 17\% and 100\% for the overall scores, respectively. The pretest shows that the student struggles and eventually abandons the problem, but in the posttest, the student demonstrates proficient use of the ladder operators to solve the problem.}\label{tabIncentiveExampleI4}
\end{figure*}

\end{document}